\documentclass[11pt]{article} 

\usepackage[utf8]{inputenc}
\usepackage[T1]{fontenc}
\usepackage[english]{babel} 
\usepackage{amsmath}
\usepackage{amsfonts}
\usepackage{amssymb}
\usepackage{amsthm}
\usepackage{euscript}
\usepackage{tikz}
\usepackage{url}
\usepackage{algorithm,algorithmic}
\usepackage{authblk}
\usepackage{a4wide}

\newtheorem{theo}{Theorem}
\newtheorem{theorem}[theo]{Theorem}

\newtheorem{lemma}[theo]{Lemma}

\newcommand{\C}{\mathbb{C}}
\newcommand{\Z}{\mathbb{Z}}

\newcommand{\HH}{\mathbf H}

\newcommand{\h}{\mathcal{H}}

\newcommand{\Prob}{\mathbb{P}}

\newcommand{\Gg}{{\mathcal G}}
\newcommand{\Hg}{{\mathcal H}}
\newcommand{\Kg}{{\mathcal K}}
\newcommand{\Vg}{{\mathcal V}}
\newcommand{\Eg}{{\mathcal E}}
\newcommand{\Fg}{{\mathcal F}}

\newcommand{\eqdef}{\stackrel{\text{def}}{=}}

\newcommand\mc[1]{\mathcal{#1}}

\begin{document}

\title{A decoding algorithm for CSS codes using the X/Z correlations}

\author[*]{Nicolas Delfosse}
\author[**]{Jean-Pierre Tillich}
\affil[*]{D\'epartement de Physique, Universit\'e de Sherbrooke, Sherbrooke, Qu\'ebec, J1K 2R1, Canada\\
	nicolas.delfosse@usherbrooke.ca}
\affil[**]{INRIA, Project-Team SECRET, 78153 Le Chesnay Cedex, France\\
	jean-pierre.tillich@inria.fr}

\renewcommand\Authands{ and }

\maketitle

\begin{abstract}

We propose a simple decoding algorithm for CSS codes taking into account the correlations between the $X$ part and the $Z$ part of the error. Applying this idea to surface codes, we derive an improved version of the perfect matching decoding algorithm which uses these $X/Z$ correlations.

\end{abstract}

\section{Introduction}
\label{section:introduction}

Low Density Parity--Check (LDPC) codes are linear codes defined by low weight parity-check equations. 
It is one of the most satisfying construction of error-correcting codes since they are both capacity approaching and endowed with an efficient decoding algorithm. It is therefore natural to investigate their quantum generalization.

Besides their use for quantum communication, quantum LDPC codes could play a central role in quantum computing. A striking difference between classical and quantum information is the fact that every manipulation of quantum bits (qubits) is very noisy. Quantum gates must therefore be implemented in a fault-tolerant way. This is realized by applying operations 
%only 
on qubits encoded by
%using 
a quantum error-correcting code. These qubits can then be regularly corrected. Some recent work of Gottesman \cite{Go13:LDPC} has shown that quantum LDPC codes are well-suited for fault-tolerance. These codes, which are defined by low weight constraints on qubits, naturally limit the propagation of errors.

The first difficulty in the generalization of LDPC codes is that most of the constructions have bounded distance (see \cite{TZ09} and references therein).  
The rare families of quantum LDPC codes equipped with a growing distance are derived from Kitaev's construction. 
Kitaev's toric code is defined by local interactions between qubits placed on a square tiling of the torus. 
Similar constructions were proposed, based on tilings of surfaces \cite{BM07a, Ze09}, 3-colored tilings \cite{BM06, De13}, 
Cayley graphs \cite{CDZ13} or other geometrical objects \cite{TZ09, KP13, FH13, Au13}.

The belief propagation decoding algorithm is an essential ingredient of the success of LDPC codes. Unfortunately, 
it is much less effective in the quantum setting due to two facts (i) the unavoidable
presence of $4$-cyles in the Tanner graph \cite{COT05a} and (ii)  the low weight generators 
can be considered as low-weight errors which are not detected by the belief propagation decoder but 
which are harmful for its convergence \cite{PC08}. 
%its  generalization is far from clear due to the unavoidable presence of 4-cycles in the Tanner graph of CSS codes \cite{PC08}.
To circumvent this obstacle, some techniques originating from classical coding theory 
were imported in quantum information recently \cite{KHIS11a, AMT12:SC}.
% such as the use of larger alphabet or spacially coupled constructions
Another direction to avoid the 4-cycles, is to consider the error, which is a quaternary vector, as a pair of binary vectors. These two binary vectors can then be decoded separately. The main problem of this point of view is that it does not consider the correlations between the two binary components of the error. In this work, we present a simple and general strategy to take into account these correlations. To illustrate this idea, we focus on surface codes equipped with the perfect matching decoding algorithm. This algorithm is usually unable to consider the correlations. Applying our method to a family of surface codes constructed from triangular tilings of a torus, we observe a clear improvement of the performance of the decoding algorithm. The depolarizing error threshold of these triangular codes is approximately $13.3\%$ while it is close to $9.9\%$ without considering the correlations.

This article is organised as follows. The definition of surface codes and the geometrical description of errors and syndrome over these codes are recalled in Section~\ref{section:surface_codes}.
Section~\ref{section:decoding} explains how decoding can be performed by using the aforementioned correlations. Section~\ref{section:PMA} is devoted to the description of the perfect matching decoding algorithm and its improvement to take into account the correlations.
%Finally, we propose a heuristic analysis of the threshold obtained by using the correlations in Section~\ref{section:threshold}.

\section{Definitions and basic properties}
\label{section:surface_codes}

{\em Error model.}
We deal here with the \emph{depolarizing channel} model which is one of the most natural quantum error model
and the quantum analog of the binary symmetric channel. Over the depolarizing channel of probability $p$, each qubit is subjected, independently, to an error $X, Y$ or $Z$ with probability $p/3$ or is left unchanged with probability $1-p$ 
where $X,Y$and $Z$ denote the usual Pauli matrices. 
An error $E$ over $n$ qubits is therefore a tensor product $\otimes_{i=1}^n E_i$ where $E_i \in \{I, X, Y, Z\}$. Errors are considered up to the phase $\{\pm 1, \pm i \}$, since quantum states are defined up to a phase.

{\em Stabilizer and CSS codes.}
A quantum code is a subspace of dimension $2^k$ of $(\C^2)^{\otimes n}$. This code encodes $k$ qubits into $n$ qubits. 
A very useful way of constructing such codes is through the stabilizer code construction \cite{Go97} where the code is
 described by the set of fixed states of a family of commuting Pauli operators $\{S_1,\dots,S_r\}$. 
In other words, the $S_i$'s are generators of the stabilizer group of the quantum code.
A particular case of this construction is the \emph{CSS construction} due to Calderbank, Shor \cite{CS96} and Steane \cite{St96}.
It consists in choosing some of these Pauli operators in $\{I, X\}^{\otimes n}$ and the rest of them in 
$\{I, Z\}^{\otimes n}$. This brings several benefits, first it simplifies the commutation relations and helps in constructing such codes
and second decoding of such codes can be achieved by decoding two binary codes as will be explained in the next paragraph.

{\em Syndrome measurement and decoding of CSS codes.}
For a stabilizer code with stabilizer generators
$\{S_1,\dots,S_r\}$ subjected to a Pauli error $E$,  it is possible to perform a measurement  which reveals 
the vector $s(E) \eqdef (E \star S_i)_{1 \leq i \leq r}$ where
$E \star S_i$ is equal to $0$ if $E$ commutes with $S_i$ and is equal to $1$ otherwise. In the case of a CSS code, the syndrome 
splits into two parts, one corresponding to the commutation with the generators belonging to $\{I,X\}^{\otimes n}$ 
and the
other one corresponding to the commutation relations with the generators in $\{I,Z\}^{\otimes n}$. Moreover, if we decompose 
the error $E$ as $E = E_X E_Z$ where $E_X \in \{I, X\}^{\otimes n}$ and $E_Z \in \{I, Z\}^{\otimes n}$ and if we let
$S_1,\dots,S_{r_X}$ be the generators which are in $\{I,X\}^{\otimes n}$ and 
$S_{r_X+1},\dots,S_{r}$ be the generators which are in $\{I,Z\}^{\otimes n}$, then the syndrome part $s_X$ which corresponds to
the generators in $\{I,X\}^{\otimes n}$ verifies 
$s_X \eqdef (E \star S_i)_{1 \leq i \leq r_X} = (E_Z \star S_i)_{1 \leq i \leq r_X}$
whereas the syndrome part $s_Z$ which corresponds to
the generators in $\{I,Z\}^{\otimes n}$ verifies 
$s_Z \eqdef (E \star S_i)_{r_X+1 \leq i \leq r} = (E_X \star S_i)_{r_X+1 \leq i \leq r}$.

Notice that if we bring in the binary matrices $\HH_X$ and $\HH_Z$ whose rows are formed for $\HH_X$, respectively  $\HH_Z$, by the  generating elements belonging to
$\{I,X\}^{\otimes n}$, respectively $\{I,Z\}^{\otimes n}$ (and replacing $I$ by $0$ and $X$ by $1$, respectively  replacing  $I$ with $0$ and $Z$ with $1$), then 
$s_X$ is nothing but the syndrome $\HH_X e_Z^T$ of the binary error $e_Z$ (obtained from $E_Z$ 
by replacing $I$ by $0$ and $Z$ by $1$), whereas
$s_Z$ is nothing but the syndrome $\HH_X e_Z^T$ of the binary error $e_Z$ (obtained from $E_Z$ 
by replacing $I$ by $0$ and $Z$ by $1$). 
In other words decoding a CSS code amounts to decode two binary codes.
This is how decoding a CSS code is usually performed. We call this decoding technique the {\em standard CSS decoder}.

{\em Tiling of a surface.}
A surface code is a CSS code associated with a tiling of surface. Let us recall the definition of a tiling of surface.
%A \emph{graph} is a pair $G=(V, E)$ such that $V$ is a set and $E$ is composed of pairs of elements of $V$. The elements of $V$ are called \emph{vertices} and the elements of $E$ are called \emph{edges}.
%JP : j'ai change la police de G,V,E pour que l'on puisse ecrire E pour l'erreur et que 
%cela ne porte pas a confusion.
A \emph{tiling of surface} is defined to be a cellular embedding of a graph $\Gg=(\Vg, \Eg)$ in a 2-manifold, that is, a surface. Without loss of generality, we can assume that the surface is smooth. We assume that the graph $\Gg$ contains neither loops nor multiple edges. This embedding defines a set of faces $\Fg$. Each face is described by the set of edges on its boundary. This tiling of surface is denoted $\Gg=(\Vg, \Eg, \Fg)$. The \emph{dual graph} of $\Gg$ is the graph $\Gg^* = (\Vg^*, \Eg^*)$ of vertex set $\Vg^*=\Fg$ such that two vertices are linked by an edge if and only if the two corresponding faces of $\Gg$ share an edge. There is a clear bijection between the edges of $\Gg$ and the edges of $\Gg^*$. This graph $\Gg^*$ is endowed with a structure of tiling of surface and its faces correspond to the vertices of $\Gg$:
%More precisely, the set of edges of $\Gg$ incident to a vertex $v \in \Vg$ induces a face of the dual graph.
these faces are induced by the set of edges of $\Gg$ incident to a vertex $v \in \Vg$.  

\par
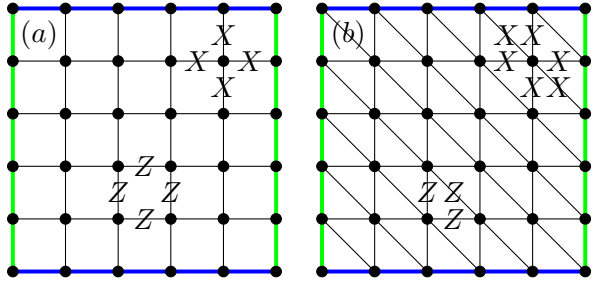
\begin{figure}
\centering
\begin{tikzpicture}[scale=.7]
\tikzstyle{every node}=[circle, draw, fill, inner sep=0pt, minimum width=4pt];

\draw[step=1cm] (0,0) grid (5,5);

\draw[color=blue, line width=1.5pt]
	(0,0)--(5,0)
	(0,5)--(5,5);

\draw[color=green, line width=1.5pt]
	(0,0)--(0,5)
	(5,0)--(5,5);

\foreach \x in {0,1,...,5}
	\foreach \y in {0,1,...,5}
		\draw (\x, \y) node {};

\tikzstyle{every node}=[inner sep=0pt, minimum width=4pt];
\node at (0.5,4.5) {$(a)$}; 
\node at (4.5,4) {$X$}; 
\node at (4,4.5) {$X$};
\node at (3.5,4) {$X$};
\node at (4,3.5) {$X$};

\node at (2.5,1) {$Z$};
\node at (3,1.5) {$Z$};
\node at (2.5,2) {$Z$};
\node at (2,1.5) {$Z$};
\end{tikzpicture}
\hspace{.2cm}
\begin{tikzpicture}[scale=.7]
\tikzstyle{every node}=[circle, draw, fill, inner sep=0pt, minimum width=4pt];

\draw[step=1cm] (0,0) grid (5,5);

\draw[color=blue, line width=1.5pt]
	(0,0)--(5,0)
	(0,5)--(5,5);

\draw[color=green, line width=1.5pt]
	(0,0)--(0,5)
	(5,0)--(5,5);

\foreach \x in {0,1,...,5}
	\foreach \y in {0,1,...,5}
		\draw (\x, \y) node {};

\foreach \x in {1,2,...,5}
	\foreach \y in {0,1,...,4}
		\draw (\x, \y)--(\x-1,\y+1);

\tikzstyle{every node}=[inner sep=0pt, minimum width=4pt];
\node at (0.5,4.5) {$(b)$}; 
\node at (4.5,4) {$X$}; 
\node at (4,4.5) {$X$};
\node at (3.5,4) {$X$};
\node at (4,3.5) {$X$};
\node at (4.5,3.5) {$X$};
\node at (3.5,4.5) {$X$};

\node at (2.5,1) {$Z$};
\node at (2,1.5) {$Z$};
\node at (2.5,1.5) {$Z$};
\end{tikzpicture}
\caption{A plaquette operator and a site operator acting on a square tiling of the torus in Fig.~$(a)$ and on a triangular tiling of the torus in Fig.~$(b)$. The opposite boundaries are identified.}
\label{fig:Kitaev}
\end{figure}

{\em Surface Codes.}
Surface codes are a special case of CSS codes. They have been introduced by Kitaev \cite{Ki97}.
% and generalized by Bombin and Martin-Delgado \cite{BM07a}. 
Assume that qubits are placed on the edges of a tiling of surface $\Gg=(\Vg, \Eg, \Fg)$. The space of the system is $\bigotimes_{e \in \Eg} \h_e$, with $\h_e = \C^2$ for every edge $e \in \Eg$. The Pauli operators acting on this space are the tensor products $\otimes_{e \in \Eg} P_e$ such that $P_e \in \{I, X, Y, Z\}$. For every edge $i \in \Eg$, denote by $X_i = \otimes_e P_e$ the Pauli operator which is the identity on every edge except on  edge $i$, where $P_i=X$. The operators $Z_i$ are defined similarly for all $i \in \Eg$.
The site operators $X_v$ and the plaquette operators $Z_f$ are the Pauli operators defined by
$$
X_v = \prod_{v \in e} X_e \quad \text{ and } \quad Z_f = \prod_{e \in f} Z_e,
$$
for every vertex $v \in \Vg$ and for every face $f \in \Fg$.
Then, the \emph{surface code} associated with the tiling of surface $\Gg$ is the CSS code fixed by the site operators and the plaquette operators. The commutation between these operators follows from the structure of the tiling. 
Note that $\HH_X$ is in this case the incidence matrix of the graph $\Gg$ and $\HH_Z$ the incidence matrix of its dual $\Gg^*$.
The site operators and the plaquette operators of Kitaev's toric codes are represented in Figure~\ref{fig:Kitaev}~$(a)$.
% This code is defined from a square tiling of the torus.

\par{\em Syndrome of a surface code.}
%\subsection{Error model and Syndrome}
% Let us recall the graphical description of the error and its syndrome.
In the case of a surface code, the syndrome has a graphical interpretation that we recall now.
Consider the surface code associated with a tiling of surface $\Gg=(\Vg, \Eg, \Fg)$.
%In what follows, we describe graphically the component $E_Z$ of the error and its syndrome.
Assume that an error $E_Z$ acts on a path $\gamma \subset \Eg$ of $\Gg$. In other words, we have $E_Z = \prod_{e \in \gamma} Z_e$. Then, the syndrome $s_X = (E_Z \star X_v)_{v \in \Vg}$ of this error is indexed by the vertices of the graph and is non-trivial if and only if the vertex $v$ is an end-point of the path $\gamma$. This follows from the fact that $E_Z$ commutes with all the operators $X_v$, except the two operators centered on the end-points of $\gamma$.
More generally, the support of the error $E_Z$ can be decomposed as a union of disjoint paths and its syndrome indicates the end-points of the support of $E_Z$.
In what follows, we denote $\partial(U) \subset \Vg$ the set of end-points of a set $U \subset \Eg$.
%The result of this measurement is represented by the syndrome vector $s_Z \in \F_2^{|V|}$. It is the indicator vector of the end-points of $E_Z$, its $i$-th component is 1 if and only if the $i$-th vertex of $G$ is an end-point of $E_Z$.

To obtain an analogous description of the error $E_X$ and its syndrome, replace the graph by its dual. Indeed, this transformation exchanges the roles of $X$ and $Z$ in the definition of the code.

The following well known lemma summarizes the graphical description of the error.
\begin{lemma} \label{lemma:surface_error}
Let $\Gg=(\Vg,\Eg,\Fg)$ be a tiling of surface and let $\Gg^*=(\Vg^*, \Eg^*, \Fg^*)$ be its dual.
An error acting on the surface code associated with $\Gg$ corresponds to a pair $(E_X, E_Z)$ such 
that $E_X \subset \Eg^*$ and $E_Z \subset \Eg$, and its syndrome is the pair $(s_X, s_Z)$ 
such that $s_X \subset \Vg$ is the set $\partial(E_Z)$ of end-points of 
$E_Z$ and $s_Z \subset \Vg^*$ is the set $\partial(E_X)$ of end-points of $E_X$.
\end{lemma}
\section{Decoding by using correlations between errors in $X$ and $Z$}
\label{section:decoding}

Virtually all decoders of  CSS codes try to recover the $E_X$ and $E_Z$ part of the error independently
by decoding two binary codes as explained in Section \ref{section:surface_codes}. There is however some
correlation between the $X$ part of the error and the $Z$ part as shown by the 
following conditional probabilities computed for a single error $E=E_X E_Z$ generated by the depolarizing channel of 
depolarizing probability $p$:
\begin{align}
&\Prob(E_Z = I | E_X = X) = 1/2\\
&\Prob(E_Z = Z | E_X = X) = 1/2
\end{align}
whereas
\begin{align}
&\Prob(E_Z = I | E_X = I) = \frac{1-p}{1-2p/3}\\
&\Prob(E_Z = Z | E_X = I) = \frac{p/3}{1-2p/3}\cdot
\end{align}
When $E_X = X$, we recognize an erasure channel and in the second case this corresponds to a binary symmetric 
channel of probability $p'' \eqdef \frac{p/3}{1-2p/3}$. This can be exploited by the following strategy for decoding. 
First, decode the $X$ component of the error. Then, erase the coefficients of $E_Z$ corresponding to the $X$ errors. Finally, decode the $Z$ component of the error, which is subjected to a combination of errors and erasures. We call such a decoder a {\em CSS decoder using $X/Z$ correlations}.

It is insightful to calculate the capacity of the two classical channels that both decoders face. The $X$ decoder
has to work for a binary symmetric channel of
crossover probability $p' \eqdef 2p/3$ whereas the $Z$ decoder has to work for 
a binary error and erasure channel, where a bit gets erased with erased with probability $p'$ and corrupted
with probability $(1-p')p''$.
The capacity of the first channel is equal to $ 1-h(p')$ whereas the capacity of the 
second channel is equal to $(1-p')(1-h(p''))$. It can be readily observed that the second capacity is always 
larger than the first one. 

This suggests two things
\begin{itemize}
\item[(i)]
 if the two binary codes have the same rate (that is if the
number of $X$ generators is the same as the number of $Z$ generators), then we may expect
that the second decoder behaves much better than the first decoder and that the probability of the whole
decoding is essentially the probability that the first decoder fails instead of being
essentially twice this probability as is usually the case for the standard CSS decoder 
described in the previous section. 
\item[(ii)] In order to fully use this decoder, the best strategy for choosing the CSS code (without using the
possible degeneracy of the code) is to choose an asymmetric CSS code where the number of $Z$ generators of the
CSS code is chosen such that the binary code associated to $\HH_Z$ has rate slightly below
$1-h(p')$ whereas the $X$ generators are chosen such that the rate of the binary code associated to
$\HH_X$ has rate slightly below $(1-p')(1-h(p''))$. This strategy of decoding 
is able to reach the hashing bound, which is equal to $1 + p \log \frac{p}{3} + (1-p) \log (1-p)$ for a depolarizing channel as explained by the following theorem.
\end{itemize}

\begin{theorem}
For any $\epsilon >0$, there exists a family of CSS codes of quantum rate $ \leq 1 + p \log \frac{p}{3} + (1-p) \log (1-p) - \epsilon$ 
for which the error probability after decoding with the CSS decoder using $X/Z$ correlations goes to $0$ as the length goes to infinity.
\end{theorem}
This theorem is proved by random coding techniques and will be given in the full version of this paper. Notice that the hashing bound is significantly bigger than 
$1-2h(p')$ which is the biggest quantum rate that random CSS codes may have in order to be decoded succesfully by the standard CSS decoder.

\section{Improvement of the Perfect Matching Decoding}
\label{section:PMA}

In this section, we recall the perfect matching decoding algorithm \cite{DKLP02} for surface codes, we discuss about its two main weakness and 
improve its performance by using the strategy outlined in Section \ref{section:decoding}. 

\subsection{The Perfect Matching Decoding Algorithm}

We consider that a surface code is subjected to a random error $(E_X, E_Z)$ generated by a depolarizing channel of probability $p$.
%After measurement of the syndrome, t
The goal of this algorithm is to determine a most likely error $E_Z$ which corresponds
here to an error of minimum weight (since in general we are in a situation where $p' \leq 1/2$)
which has syndrome $s_X$. The component $E_X$ is decoded similarly in the dual graph.
%The distribution of the error $E_Z$ corresponds to a binary symmetric channel of probability $p' = 2p/3$:
%$$
%\Prob(E_Z) = p'^{|E_Z|} (1-p')^{n-|E_Z|},
%$$
%where $|E_Z|$ is the weight of the error, that is, its number of non-identity coefficients.
%Therefore, when $p'<1/2$, a most likely error is exactly an error of minimum weight.

To determine an error $E_Z \subset \Eg$ of minimum weight, given its end-points $s_X=\partial(E_Z) \subset \Vg$, we are looking for a set of paths whose end-points are exactly $s_X$ and whose size is minimum. Algorithm~\ref{algo:PMA} computes such a set using Edmonds' minimum weight perfect matching algorithm \cite{Ed65a, Ed65b, Ko09}.
This decoding algorithm first computes the distance graph associated with a syndrome $s \subset \Vg$.
It is the weighted complete graph $\Kg(s)$, with vertex set $s=\{s_1, s_2, \dots, s_m\}$, such that the weight of the edge $\{s_i, s_j\}$ is the distance $d(s_i, s_j)$ in 
%the graph 
$\Gg$.
The second step of the algorithm is the determination of a minimum weight perfect matching $M$
%$M \subset \Eg(\Kg(s))$ 
in 
%the graph 
$\Kg(s)$. Recall that a perfect matching in a graph $\Hg$ is a set of edges of $\Hg$ meeting all the vertices of $\Hg$ exactly once. 
With each edge $\{v_i, v_j\} \in M$, we associate a geodesic of $\Gg$ joining the vertices $v_i$ and $v_j$. 
Denote by $\mc \Gg(v_i, v_j)$ this geodesic.
The algorithm returns the symmetric difference of all the geodesics corresponding to the edges of $M$. 
It is the support of a most likely error of syndrome $s$.

\begin{algorithm}
\caption{Perfect Matching Decoding}
\label{algo:PMA}

\begin{algorithmic}[1]
\REQUIRE A graph $\Gg = (\Vg, \Eg)$, a subset $s\subset \Vg$ of odd size.
\ENSURE A subset $x \subset \Eg$ of minimum size with end-points $\partial(x) = s$.

%\STATE Compute the distance between all the pairs of vertices of $s$.
\STATE Construct the distance graph $\Kg(s)$ associated with $s$.

\STATE Determine a minimum weight perfect matching $M$
%$M \subset \Eg(\Kg(s))$.

\STATE return the symmetric difference of all the geodesics $\mc \Gg(v_i, v_j)$ for $\{v_i, v_j\} \in M$.
\end{algorithmic}
\end{algorithm}

\subsection{Degeneracy and Correlations}

We now discuss of two cases of failure of the perfect matching decoding algorithm and their effect on the performance.

First, by definition, surface codes are fixed by the plaquette operators of the tiling.
Thus two errors which differ in a sum of plaquettes (or faces) have exactly the same effect on the quantum code. We should thus look for the most likely error coset up the sums of faces instead of the most likely error. This phenomenon is called \emph{degeneracy}. The threshold of the toric code obtained by taking account optimally of the degeneracy has been estimated using an Ising model interpretation of the decoding problem \cite{DKLP02}. This threshold is close to $p=0.163$ whereas the perfect matching algorithm reaches its threshold at approximately $p=0.155$.
Note that the renormalization group approach of \cite{DP10a} is one of the rare decoding algorithm of the toric code which is able to make use of the degeneracy of the code.

The second possibility of improvement of the decoding algorithm is the most important potential gain in the performance.
It is the correlation between the 2 components, $E_X$ and $E_Z$, of the error
and consists in using the decoding strategy explained in Section \ref{section:decoding}.
%In the perfect matching decoding algorithm, we decode independently these two error vectors. But this is not coherent with the depolarizing noise.
%If the X comp of the error is X then the Z component is also Z error with proba 1/2 (when the Pauli error is Y) or is I with proba 1/2 (when the Pauli error is X).
The threshold of the toric code using the $X/Z$ correlations has been estimated close to $0.189$ with the Ising model correspondence \cite{BAOKM12} and is approximately $0.185$ with the non-efficient Metropolis decoding algorithm \cite{WL12}.

These two remarks are generally true for all surface codes.

\subsection{A Correlated Perfect Matching Algorithm}

To implement the decoding strategy of Section \ref{section:decoding} we need to be able to correct errors and erasures
when decoding the $E_Z$ part. The correction of combinations of errors and erasures for topological codes has been considered by Stace, Barrett, and Doherty \cite{BDS09}. We choose here to adapt Algorithm~\ref{algo:PMA} to find a most likely error for this error model.

\begin{figure}
\centering
\begin{tikzpicture}[scale = .6]
\tikzstyle{every node}=[circle, draw, fill, inner sep=0pt, minimum width=2pt];

\draw[step=1cm] (0,0) grid (5,5);

\draw[color=blue, line width=1.5pt]
	(0,0)--(5,0)
	(0,5)--(5,5);

\draw[color=green, line width=1.5pt]
	(0,0)--(0,5)
	(5,0)--(5,5);

\foreach \x in {0,1,...,5}
	\foreach \y in {0,1,...,5}
		\draw (\x, \y) node {};

\tikzstyle{every node}=[inner sep=0pt, minimum width=4pt];
\node at (0.5,4.5) {$(a)$}; 
\node at (2.5,4) {$Z$}; 
\node at (2,3.5) {$Y$}; 
\node at (1,2.5) {$X$};
\node at (3,1.5) {$Y$}; 
\node at (3,0.5) {$Z$}; 
\end{tikzpicture}
\hspace{.5cm}
\begin{tikzpicture}[scale = .6]
\tikzstyle{every node}=[circle, draw, fill, inner sep=0pt, minimum width=2pt];

\draw[step=1cm] (0,0) grid (5,5);

\draw[color=blue, line width=1.5pt]
	(0,0)--(5,0)
	(0,5)--(5,5);

\draw[color=green, line width=1.5pt]
	(0,0)--(0,5)
	(5,0)--(5,5);

\foreach \x in {0,1,...,5}
	\foreach \y in {0,1,...,5}
		\draw (\x, \y) node {};

\tikzstyle{every node}=[inner sep=0pt, minimum width=4pt];
\node at (0.5,4.5) {$(b)$}; 
\node at (2,3.5) {$X$}; 
\node at (1,2.5) {$X$};
\node at (3,1.5) {$X$}; 
\end{tikzpicture}

\vspace{.2cm}
\begin{tikzpicture}[scale = .6]
\tikzstyle{every node}=[circle, draw, fill, inner sep=0pt, minimum width=2pt];

\draw[step=1cm] (0,0) grid (5,5);

\draw[color=blue, line width=1.5pt]
	(0,0)--(5,0)
	(0,5)--(5,5);

\draw[color=green, line width=1.5pt]
	(0,0)--(0,5)
	(5,0)--(5,5);

\foreach \x in {0,1,...,5}
	\foreach \y in {0,1,...,5}
		\draw (\x, \y) node {};

\draw[color=red, dashed, line width=1.5pt]
	(1,2)--(1,3)
	(2,3)--(2,4)
	(3,1)--(3,2);

\tikzstyle{every node}=[inner sep=0pt, minimum width=4pt];
\node at (0.5,4.5) {$(c)$}; 
\draw
	(3, 0) node[circle, fill=red] {1}
	(3, 2) node[circle, fill=red] {1}
	(2, 3) node[circle, fill=red] {1}
	(3, 4) node[circle, fill=red] {1};	
\end{tikzpicture}
\hspace{.5cm}
\begin{tikzpicture}[scale = .6]
\tikzstyle{every node}=[circle, draw, fill, inner sep=0pt, minimum width=2pt];

\draw[step=1cm] (0,0) grid (5,5);

\draw[color=blue, line width=1.5pt]
	(0,0)--(5,0)
	(0,5)--(5,5);

\draw[color=green, line width=1.5pt]
	(0,0)--(0,5)
	(5,0)--(5,5);

\foreach \x in {0,1,...,5}
	\foreach \y in {0,1,...,5}
		\draw (\x, \y) node {};

\tikzstyle{every node}=[inner sep=0pt, minimum width=4pt];
\node at (0.5,4.5) {$(d)$}; 
\node at (3,0.5) {$Z$}; 
\node at (3,1.5) {$Z$};
\node at (2,3.5) {$Z$};
\node at (2.5,4) {$Z$};
\end{tikzpicture}

\caption{An example of error correction using Algorithm~\ref{algo:corPMA}. (a) An error for Kitaev's toric code. (b) The component $E_X$ computed at Step 1. of Algorithm~\ref{algo:corPMA}. (c) The syndrome $s_X$ of $E_Z$ is given by the vertices marked with '1'. The dashed edges form the erasure defined from $E_X$. (d) The $Z$ component estimated in Step 2. of Algorithm~\ref{algo:corPMA}. It is the an error of syndrome $s_X$ which has minimum weight on the non-erased qubits.}
\label{fig:error}
\end{figure}
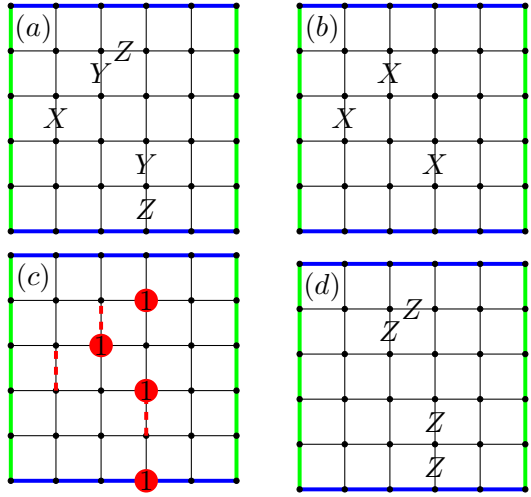

Denote by $E_Z^e$ the restriction of $E_Z$ to the erased positions, that is the positions (or edges)
%coefficients 
such that $E_X =X$, and denote by $E_Z^{\bar e}$ its restriction to the non-erased positions.
%The probability of the vector $E_Z$, given $E_X$, is
%$$
%\Prob(E_Z | E_X) = p''^{|E_Z^{\bar e}|} (1-p'')^{n-|E_X| - |E_Z^{\bar e}|} (1/2)^{|E_X|}.
%$$
%This probability does not depend on the value of $E_Z$ on the erased positions. When $p''<1/2$, it suffices to minimize the weight of $E_Z^{\bar e}$. 
To find the error $E_Z$ of syndrome $s_X$ such that the weight of $E_XE_Z$ is minimum, 
we just have to modify the distance function in Algorithm~\ref{algo:PMA}.
 We introduce the $e$-distance $d_e$, associated with an erasure $e$. 
The usual distance between two vertices of a graph $\Gg$ is the length of a shortest path joining these two vertices. 
The distance $d_e$ is defined similarly but the length of a path is its number of non-erased edges. 
An $e$-geodesic between two vertices $u$ and $v$ of $\Gg$ is a path of $\Gg$ of length $d_e(u, v)$ joining these two vertices. 
This provides us a version of the perfect matching algorithm to correct combinations of errors and erasures. 
It is presented in Algorithm~\ref{algo:ePMA}. The distance graph based on the $e$-distance $d_e$ is denoted $\Kg^e(s)$. 
The notation $\mc \Gg^e(u, v)$ represents a $e$-geodesic between $u$ and $v$.

% As we can see with the example of the metropolis algorithm, It seems difficult to find an algorithm which both benefits of the correlations and stays efficient. Our priority is to conserve the efficiency of the decoding algorithm. Therefore, we propose a simple method to consider partially the correlations in the perfect matching decoding algorithm for a small cost in complexity.

% The key ingredient is the following remark. Consider an error $E_X E_Z$, acting over 1 qubit, generated by the depolarizing channel. We have $E_X \in \{I, X\}$ and $E_Z \in \{I, Z\}$. By definition of this channel, we know that
% \begin{align}
% &\Prob(E_Z = I | E_X = X) = 1/2\\
% &\Prob(E_Z = Z | E_X = X) = 1/2
% \end{align}
% and
% \begin{align}
% \Prob(E_Z = I | E_X = I) = \frac{1-p}{1-2p/3}\\
% \Prob(E_Z = Z | E_X = I) = \frac{p/3}{1-2p/3}.
% \end{align}
% When $E_X = X$, we recognize an erasure channel and in the second case this corresponds to a binary symmetric channel of probability $p'' = \frac{p/3}{1-2p/3}$.

\begin{algorithm}
\caption{Perfect Matching Decoding for errors and erasures}
\label{algo:ePMA}

\begin{algorithmic}[1]
\REQUIRE A graph $\Gg = (\Vg, \Eg)$, a subset $s\subset \Vg$ of odd size, a set of erased edges $e \subset \Eg$.
\ENSURE A subset $x \subset \Eg$ with end-points $\partial(x) = s$ such that the cardinality of $x \backslash e$ is minimum.

%\STATE Compute the distance between all the pairs of vertices of $s$.
\STATE Construct the $e$-distance graph $\Kg^e(s)$ associated with $s$.

\STATE Determine a minimum weight perfect matching $M \subset \Eg(\Kg^e(s))$.

\STATE return the symmetric difference of all the $e$-geodesics $\mc \Gg^e(v_i, v_j)$ for $\{v_i, v_j\} \in M$.
\end{algorithmic}
\end{algorithm}

Combining Algorithm~\ref{algo:PMA} and Algorithm~\ref{algo:ePMA}, we obtain Algorithm~\ref{algo:corPMA}, which is an improved version of the Perfect Matching Decoding algorithm taking partially account of the $X/Z$ correlations.

\begin{algorithm}
\caption{Correlated Perfect Matching Decoding}
\label{algo:corPMA}

\begin{algorithmic}[1]
\REQUIRE A tiling $G = (\Vg, \Eg, \Fg)$, a syndrome $(s_X, s_Z) \subset V \times V^*$.
\ENSURE An error $(E_X, E_Z) \subset \Eg^* \times \Eg$ of syndrome $(s_Z, s_X)$, such that $|E_X|$ is minimum and $|E_XE_Z|$ is minimum given $E_X$. 

\STATE Compute $E_X$ by applying Algorithm~\ref{algo:PMA} to $s_Z$ in the dual graph $\Gg^*$.

\STATE Compute $E_Z$ by applying Algorithm~\ref{algo:ePMA} to $s_X$ and $e=E_X$ in the graph $\Gg$.

\STATE return $(E_X, E_Z)$.
\end{algorithmic}
\end{algorithm}

An example of error over the toric code that can not be corrected with the usual perfect matching decoding but that is corrected with Algorithm~\ref{algo:corPMA} is represented in Figure~\ref{fig:error}.

For Kitaev's toric codes, we obtain a slight improvement of the decoding performance using Algorithm~\ref{algo:corPMA}, but we cannot overcome the usual threshold since the $E_X$ part of the error is decoded using a standard perfect matching algorithm. Nevertheless, as explained in Section~\ref{section:decoding}, the use of the $X/Z$~correlations is well appropriate to asymmetric CSS codes. To define asymmetric surface codes, it suffices to consider non-self dual tilings.
%Recall that asymmetric codes are extremely important to understand because they are appropriate to asymmetric noise models, which are more realistic than the depolarizing channel.

\begin{figure}
\begin{center}
\includegraphics[scale=.6]{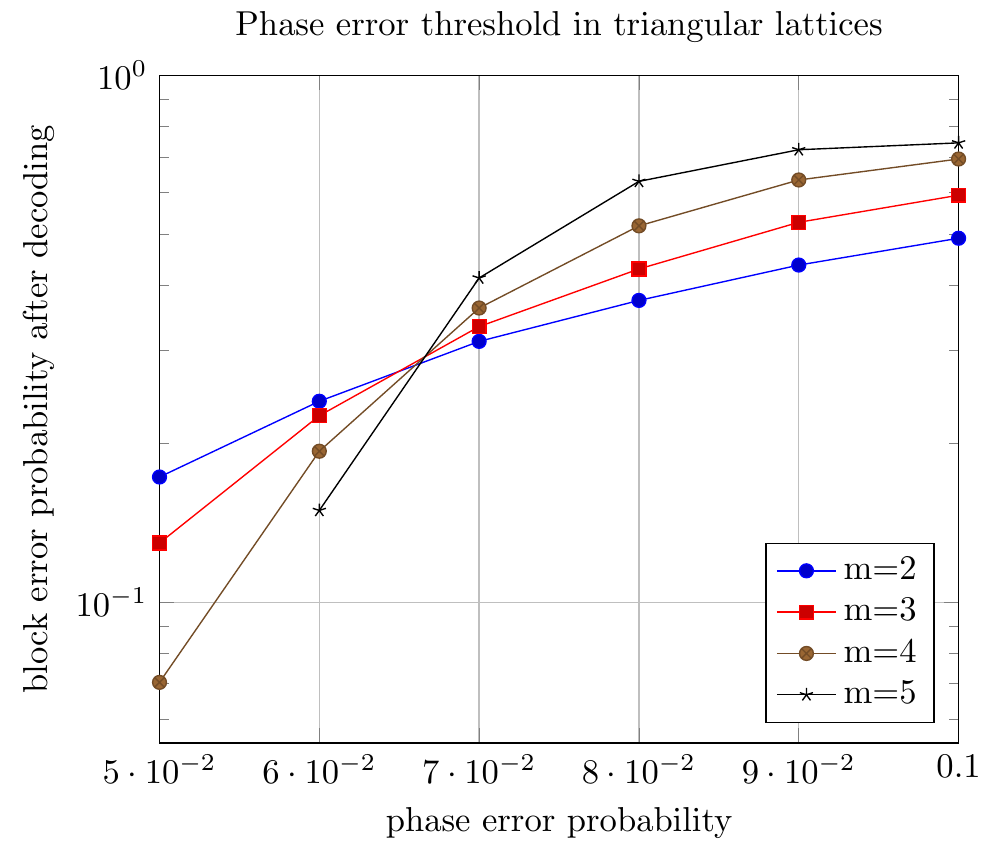}
\includegraphics[scale=.6]{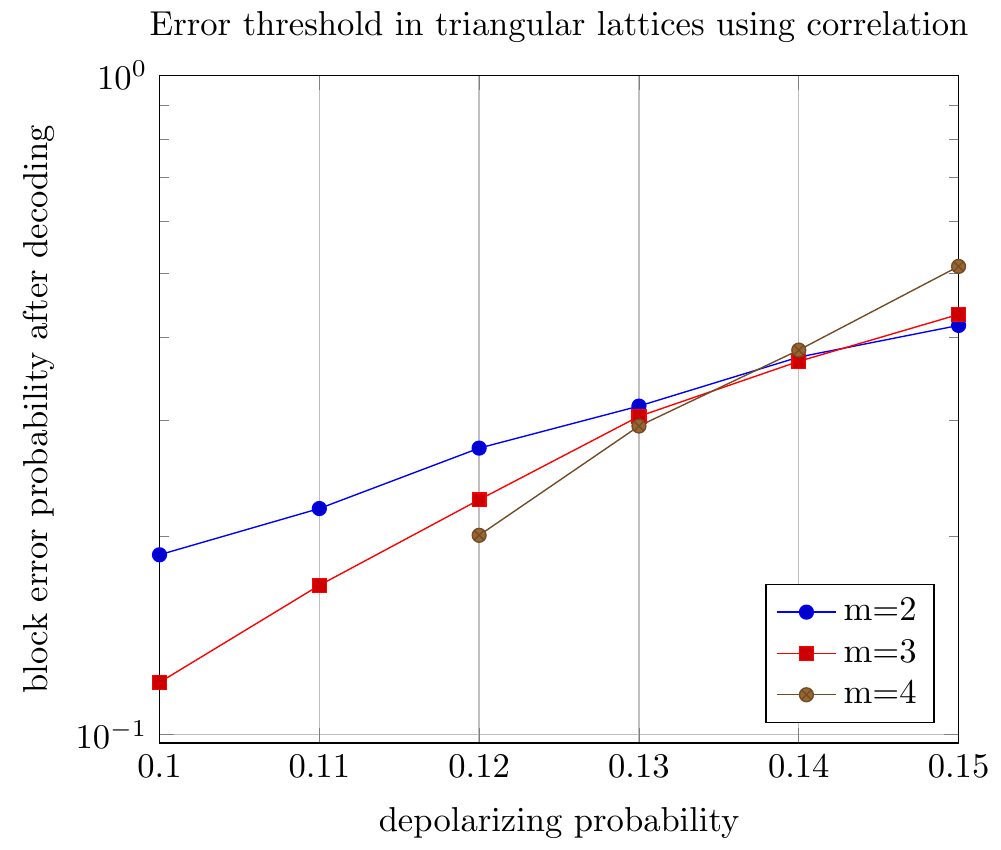}
\end{center}
\caption{Phase decoding performance of Algorithm~\ref{algo:PMA} and depolarizing decoding performance of Algorithm~\ref{algo:corPMA} for triangular toric codes of length~$3.2^{2m}$}
\label{fig:tri}
\end{figure}

A natural construction of asymmetric surface codes is the family derived from triangular lattices of the torus. For example, the Cayley graph of the group $\Z/m\Z \times \Z/m\Z$ and the generating set $\{\pm (1, 0), \pm (0, 1), \pm (1,-1)\}$, described in Figure~\ref{fig:Kitaev}~$(b)$, clearly defines a triangular tiling of the torus. Using Algorithm~\ref{algo:PMA}, we remark a threshold for the correction of phase errors at $p'=0.066$ in Figure~\ref{fig:tri}, whereas the bit-flip error threshold, observed in the dual graph (which is a
%an JP le "h" se prononce ici, pas de "an"
 hexagonal lattice), is very high (more than $p'=0.14$ for this family of tiling).
This implies a depolarizing error threshold at $p=3p'/2 = 0.099$ for the standard perfect matching algorithm, while Algorithm~\ref{algo:corPMA} leads to a depolarizing error threshold at approximately $p=0.133$. This good performance is explained by the fact that, while the phase error threshold is low, the error correction in the dual graph exhibits a very good performance and the bit-flip error threshold is high. This allows Algorithm~\ref{algo:corPMA} to take into account the $X/Z$ correlations.

%\begin{figure}
%\begin{center}
%\includegraphics[scale=.6]{tri_depo.pdf}
%\end{center}
%\caption{Depolarizing decoding performance of triangular toric codes of length $3.2^{2m}$}
%\label{fig:tri_depo}
%\end{figure}

%\input{analysis}
\section{Concluding remarks}
\label{section:conclusion}

We proposed a decoding algorithm for CSS codes partially taking into account the correlations between the $X$ component and the $Z$ component of the error for a depolarizing channel. 
Applied to triangular toric codes, this algorithm exhibits a good performance and clearly improves the threshold.
%Our method to take advantage of the $X/Z$-correlations 
%can be applied to all C to a wide class of quantum codes. For example, it can be immediately applied to hyperbolic surface codes \cite{FML02, Ze09}.
It could be applied to other classes of codes, for instance for color codes, 
where the decoding algorithm by projection onto 3 surface codes can be adapted to take into account the correlations between the 3 surface codes \cite{De14:couleur}.

\section*{Acknowledgments}

The authors wish to thank David Poulin for useful discussions. The work of both authors was supported in part by the French PEPS ICQ2013 program (TOCQ project) 
and Nicolas Delfosse was supported by the Lockheed Martin Corporation.

%\bibliographystyle{plain}
%\bibliography{biblio}

\newcommand{\SortNoop}[1]{}

\end{document}